\documentclass[onecolumn,floatfix,superscriptaddress,showpacs,showkeys,nofootinbib,preprint]{revtex4}
\usepackage{epsfig}
\usepackage{amssymb,latexsym,amsmath}
\newcommand{\eq}[1]{\begin{align} #1 \end{align}}

\begin{document}

\title{Hadron--Resonance Gas
at Freeze--out:  \\
Reminder on Importance of Repulsive Interactions
}

\author{V.V. Begun}
 \affiliation{Bogolyubov Institute for Theoretical Physics, %03680
 Kiev, Ukraine}
 \affiliation{Institut f\"{u}r Kernphysik, Goethe--Universit\"{a}t, % D-60438
 Frankfurt am Main, Germany}
\author{M. Ga\'zdzicki}
 \affiliation{Institut f\"{u}r Kernphysik, Goethe--Universit\"{a}t, %D-60438
 Frankfurt am Main, Germany}
 \affiliation{Instytut Fizyki, Jan Kochanowski University, %PL-25-406
 Kielce, Poland}

\author{M.I. Gorenstein}
 \affiliation{Bogolyubov Institute for Theoretical Physics, %03680
 Kiev, Ukraine}
 \affiliation{Frankfurt Institute for Advanced Studies, %D-60438
 Frankfurt am Main, Germany}

%\vspace*{2cm}
\begin{abstract}
An influence of the repulsive interactions on matter properties
is considered within the excluded volume van der Waals
hadron-resonance gas model.
Quantitative results are presented for matter at the chemical
freeze-out in central nucleus-nucleus collisions at relativistic
energies.
In particular, it is shown that repulsive interactions connected
to non-zero size of created particles lead to a significant
decrease of  collision energy at which the net-baryon density has
a maximum.
A position of the transition point from baryon to meson dominated matter
depends on the
difference between baryon and meson hard-core radiuses.

\end{abstract}



\maketitle

\section{Introduction}

Statistical models of the hadron gas are an important tool to extract
properties of  matter created in relativistic nucleus-nucleus
collisions (see, e.g.,
Refs.~\cite{st1,st2,st3,st3a,st3b,st4,Gaz2005}). Basic parameters
of these models are the matter temperature $T$,  baryon chemical
potential $\mu_B$, and volume $V$. If supplemented by additional
model parameters  monitoring deviations from the
chemical equilibrium~\cite{Rafelski}, they approximately fit rich
data on mean hadron multiplicities in a broad range of reactions,
from e$^+$+e$^-$, p+p, and  p+$\bar{\rm{p}}$~\cite{Becattini} at
low energies to central Pb+Pb collisions at the highest LHC
energy~\cite{LHC}.

The most popular version of the statistical models of hadron
matter is the ideal hadron-resonance gas (I--HRG), i.e., a
statistical system of non-interacting hadrons and resonances.
It is argued, based on the Dashen, Ma and Bernstein
theorem~\cite{DM}, that resonances introduced to the ideal hadron
gas take into account attractive interactions between hadrons. The
repulsive part of the interactions between hadrons is usually
accounted for by the van der Waals excluded volume procedure
generalized to the relativistic case of a variable number of
hadrons~\cite{vdw}. The resulting excluded volume model is no
longer an ideal gas model, and in this paper it will be denoted as
the EV--HRG model.
Another popular example of modelling attractive and repulsive
interactions between hadrons is the relativistic mean field theory
in a form of the Walecka model \cite{Wal} and its different
modifications (see e.g., the recent paper \cite{RMF} and
references therein). In this approach, scalar and vector meson
fields describe respectively the attractive and repulsive forces
between baryons.

Both the attractive and repulsive interactions are important for
the qualitative as well as quantitative description of the
properties of strongly interacting matter.
For example, the nucleon-nucleon potential includes both parts --
attractive at large and repulsive at small distances. The
presence of both attractive and repulsive interactions between
nucleons is crucial for the existence of stable nuclei.
Moreover, an important undesired  feature of the I--HRG model at
high temperatures was noted by the authors of
Refs.~\cite{vdwa,vdw2}. Due to the large number of different types
of baryons and mesons, the point-like hadrons would always become
the dominant phase at very high energy density. Just the excluded
volume effects ensure a transition from a gas of hadrons and
resonances to the quark-gluon plasma. Thus, one needs the EV--HRG
equation of state for hydrodynamic models of nucleus-nucleus
collisions (see, e.g.,
Refs.~\cite{Hama-2004,Werner-Karpenko-2010,Sa}). Note that the
lattice QCD calculations also indicate a presence of excluded
volume corrections~\cite{HRG-vs-Lattice}.

The aim of this paper is to recall the role of repulsive
interactions between hadrons. It is shown that the excluded volume
hadron-resonance gas model yields  different properties of matter
at the freeze-out  than the ideal hadron-resonance gas, if
densities and their collision energy dependence are considered.
Quantitative results are presented for two examples, namely
collision energy dependence of the  net-baryon
density~\cite{RC2006} as well as the ratio of baryon and meson
entropy density~\cite{s-bar,Cleymans2005,Cleymans2011}. They are
selected, because of conjectures that the  maximum of
net-baryon density as well as the transition between baryon and
meson dominated matter may be related to the onset of
deconfinement observed in central Pb+Pb collisions at the CERN SPS
energies~\cite{Horn}. Note that chemical freeze-out parameters $T$
and $\mu_B$ in nucleus-nucleus collisions are straightforwardly
connected to the data on hadron multiplicities. The energy range
considered in this paper is presently studied experimentally at
the CERN SPS~\cite{NA61} and the BNL RHIC~\cite{BES}. In future
this effort will be extended by experiments at new accelerators,
at the JINR NICA~\cite{NICA} and the FAIR
SIS--100~\cite{CBM,FAIR,Galatyuk}.

The paper is organized as follows. In Sec.~\ref{sec-id} the ideal
hadron-resonance gas model is introduced. The excluded volume
hadron-resonance gas model is presented in Sec.~\ref{sec-vdw},
where also quantitative results for densities at the chemical
freeze-out in central Pb+Pb collisions are given and discussed. A
summary given in Sec.~\ref{sec-sum} closes the paper.

\section{Ideal Hadron--Resonance Gas}\label{sec-id}

In the grand canonical ensemble the pressure of the I--HRG is
given by
\eq{\label{pres-id}
 p^{id}
 %(T,\mu_1,\ldots)
 ~=~\sum_i p^{id}_i(T,\mu_i)
 %\nonumber \\
~=~ \sum_i
\frac{d_i}{6\pi^2}\int_0^{\infty}\frac{k^4\,dk}{(k^2+m_i^2)^{1/2}}~
\left[ \exp\left(\frac{\sqrt{k^2+m_i^2}-\mu_i}{T}\right)~+~\eta
\right]^{-1}~,
}
where $T$ is the system temperature, $\eta=-1$ and $\eta=1$ for
bosons and fermions, respectively, while $\eta=0$ corresponds to
the Boltzmann approximation. For a hadron  $i$, $m_i$ is its mass
and $d_i$ is the  spin degeneracy. The chemical potential  is
given by
\eq{\label{mui}
\mu_i~=~b_i\mu_B~+~s_i\mu_S~+~q_i\mu_Q
}
with
%
%\eq{\label{qi}
%
$b_i=0,\pm1, ~s_i=0,\mp 1,\mp 2, \mp 3$ and $q_i=0,\pm 1,\pm 2$~
%
%}
%
for hadrons.
The number density of a hadron $i$ reads:
\eq{\label{ni-id}
 n_i^{id}(T,\mu_i)~=~T\,\frac{\partial p^{id}}{\partial
\mu_i}~=~ \frac{d_i}{2\pi^2}\int_0^{\infty}k^2\,dk~ \left[
\exp\left(\frac{\sqrt{k^2+m_i^2}-\mu_i}{T}\right)~\pm~1
\right]^{-1}~.
}
Considering the temperature $T$, baryon chemical potential
$\mu_B$, and volume $V$ as free parameters one can fit mean hadron
multiplicities $\langle N_i\rangle=Vn_i$ measured in relativistic
nucleus-nucleus collisions at each collision energy. In this
analysis, $\mu_S$ and $\mu_Q$ are expressed as functions of $T$
and $\mu_B$ when the conditions on strangeness, $\langle
S\rangle=0$, and electric to baryon charge ratio, $\langle
Q\rangle/\langle B\rangle = Z/A$, are taken into account. Most of
experimental data on nucleus-nucleus collisions concern yields of
long-lived hadrons, which include products of resonance decays.
This requires a proper treatment of short-lived resonances, namely
the products of their strong and electromagnetic decays should be
added to the mean multiplicities of stable hadrons. In this paper
the numerical implementation of the hadron-resonance gas model
provided by the THERMUS package~\cite{THERMUS} is used to
calculate the relevant quantities according to
Eqs.~(\ref{T}-\ref{gamma-S}). Particles and resonances [all mesons
up to $K^*_4$(2045)] and baryons (up to $\Omega^-$), quantum
statistics, as well as the width of resonances are included.

\begin{figure}[ht!]
\begin{center}
 \epsfig{file=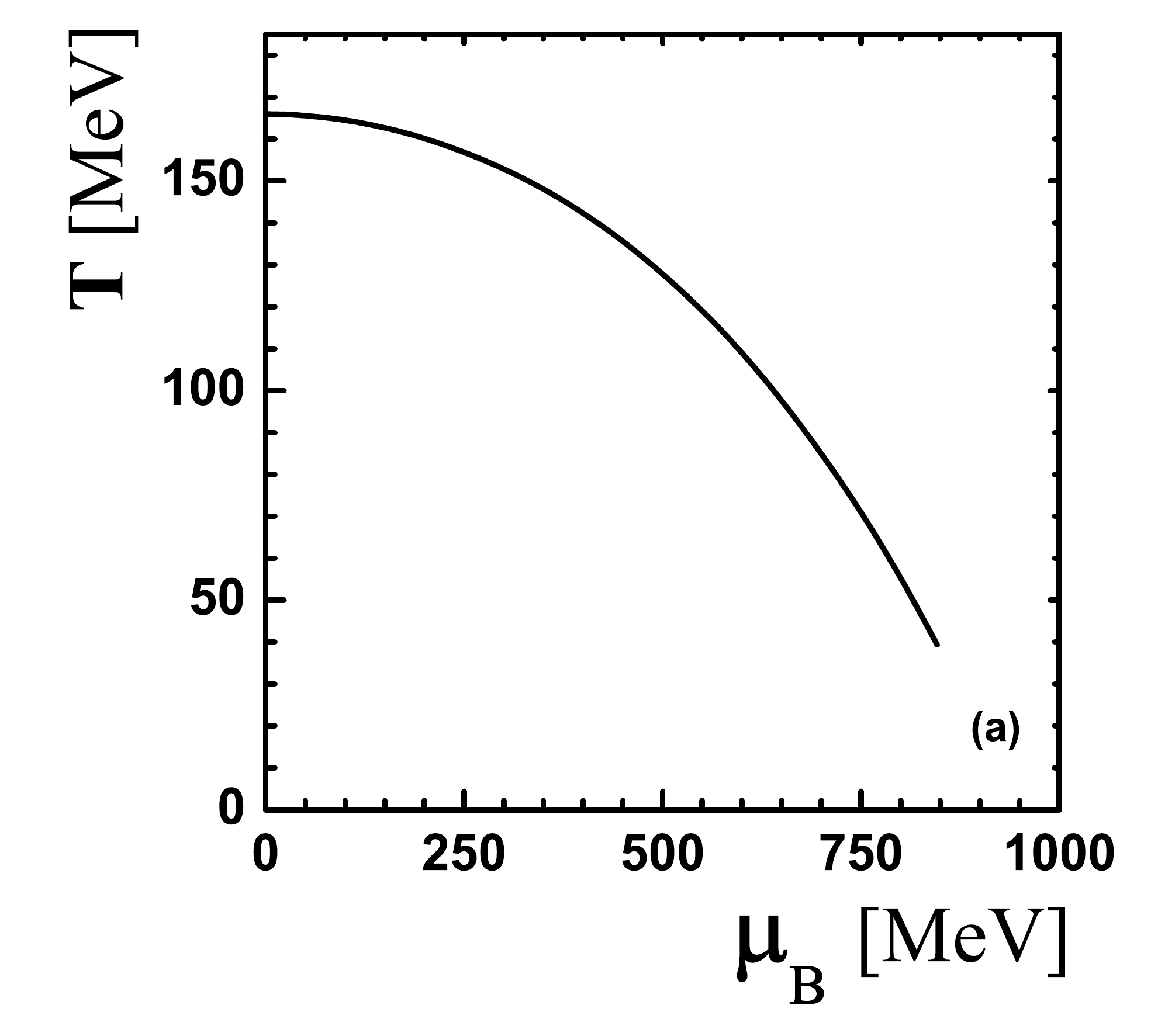,width=0.462\textwidth}~~
 \epsfig{file=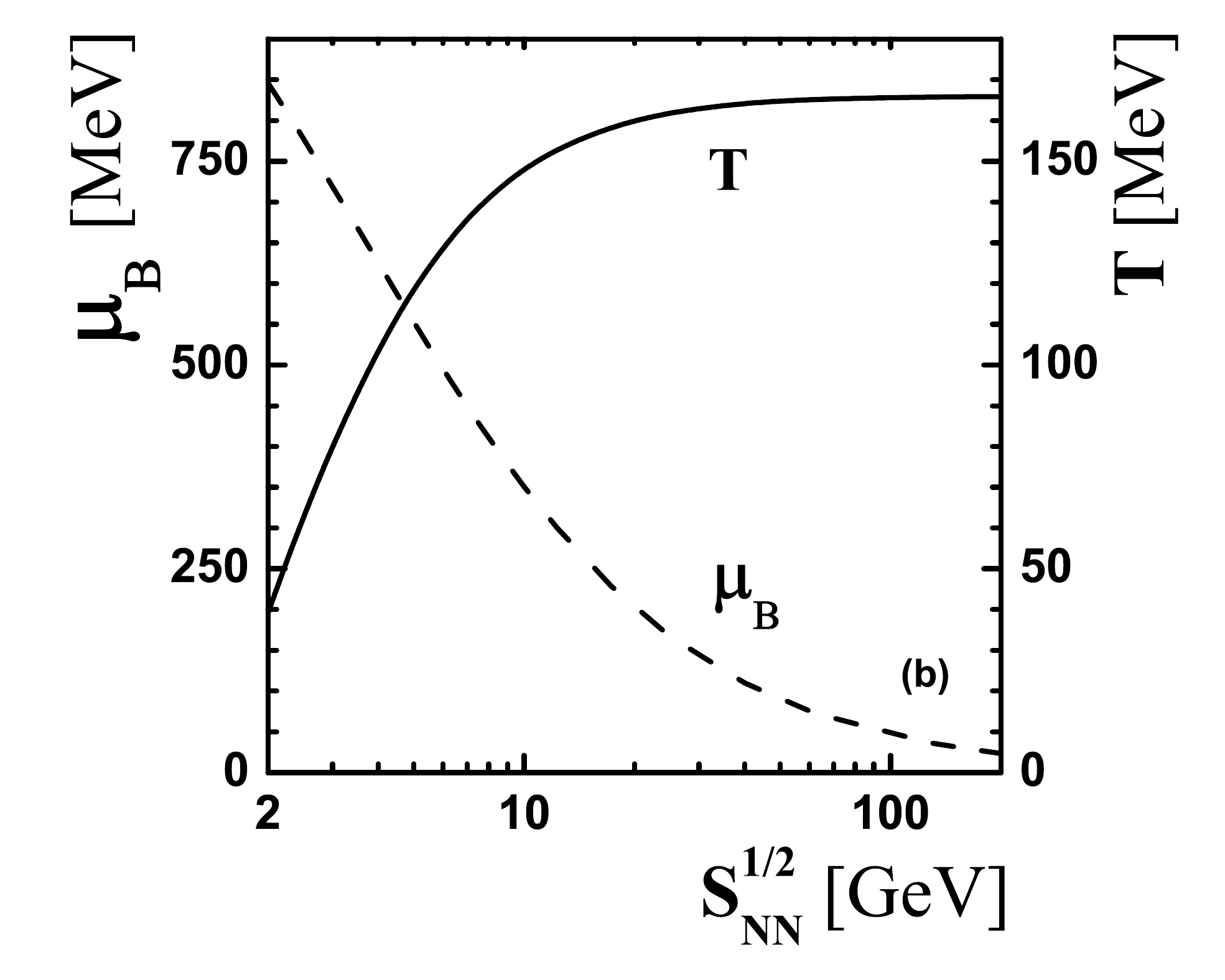,width=0.528\textwidth}
 \caption{(Color online) (a): The chemical freeze-out line $T=T(\mu_B)$. (b): The $T$
 and $\mu_B$ along the chemical freeze-out as a function
 of $\sqrt{s_{NN}}$.}\label{fig-T}
 \end{center}
\end{figure}

The analysis of central Pb+Pb (Au+Au) collisions registered by
experiments at SIS, AGS, SPS, and RHIC allows to establish
the collision energy dependence of $T$ and $\mu_B$ which can be
parameterized as~\cite{Cleymans2005}:
\eq{\label{T}
 T ~&=~ 0.166\,\text{GeV} ~-~ 0.139\,\text{GeV}^{-1}\mu_B^2 ~-~ 0.053\,\text{GeV}^{-3} \mu_B^4~,
% \nonumber
 \\
 \mu_B ~&=~ \frac{1.308\,\text{GeV}}{1 ~+~ 0.273\,\text{GeV}^{-1}\,\sqrt{s_{NN}}}~,
 }
where $\sqrt{s_{NN}}$ is the center-of-mass energy of a nucleon
pair. The chemical freeze-out line, $T=T(\mu_B)$, as well as the
energy dependence of the $T$ and $\mu_B$ parameters are shown in
Figs.~\ref{fig-T} (a) and (b), respectively.
The net-baryon density $\rho_B$, entropy density $s$, and energy
density $\varepsilon$ can be found from the system pressure $p$
using the thermodynamical relations:
\eq{\label{rhoB}
\rho_B~=~\frac{\partial p}{\partial \mu_B}~,~~~~~~
s~=~\frac{\partial p}{\partial T}~,~~~~~~
\varepsilon~=~T\frac{\partial p}{\partial T} ~+~\mu \frac{\partial
p}{\partial \mu}~-~p~.
%s~=~\frac{\partial p}{\partial T}~.
%
}
With the chemical freeze-out parameters given by Eq.~(\ref{T}) and
the ideal gas expression Eq.~(\ref{pres-id}) for the system
pressure, one finds the quantities in Eq.~(\ref{rhoB}) as
functions of the collision energy.
The I--HRG model is based on the assumption of complete thermal
and chemical equilibrium. An additional I--HRG parameter, the
strangeness suppression factor $\gamma_S$, has to be introduced to
account for deviations of strange hadron multiplicities from
chemical equilibrium~\cite{Rafelski}. Its dependence on $T$ and
$\mu_B$ obtained by fitting  hadron yields measured in the full
phase space can be parameterized as~\cite{Gaz2005}
\eq{\label{gamma-S}
 \gamma_S~ = ~1-
 0.396\,\exp\left(\,-~1.23\,\frac{T}{\mu_B}\right)~.
 }
At the AGS and SPS energies the $\gamma_S$ parameter is
significantly smaller than $1$, which means the under-saturation
of strange hadron yields with respect to the chemical equilibrium.
The $\gamma_S$ parameter should be included in the model if a
proper description of strange hadron yields is required.
However, the relation between $T$ and $\mu_B$ as well as the
chemical freeze-out line Eq.~(\ref{T}), obtained within the models
with and without the $\gamma_S$ parameter, is similar.

\begin{figure}[ht!]
\begin{center}
 \epsfig{file=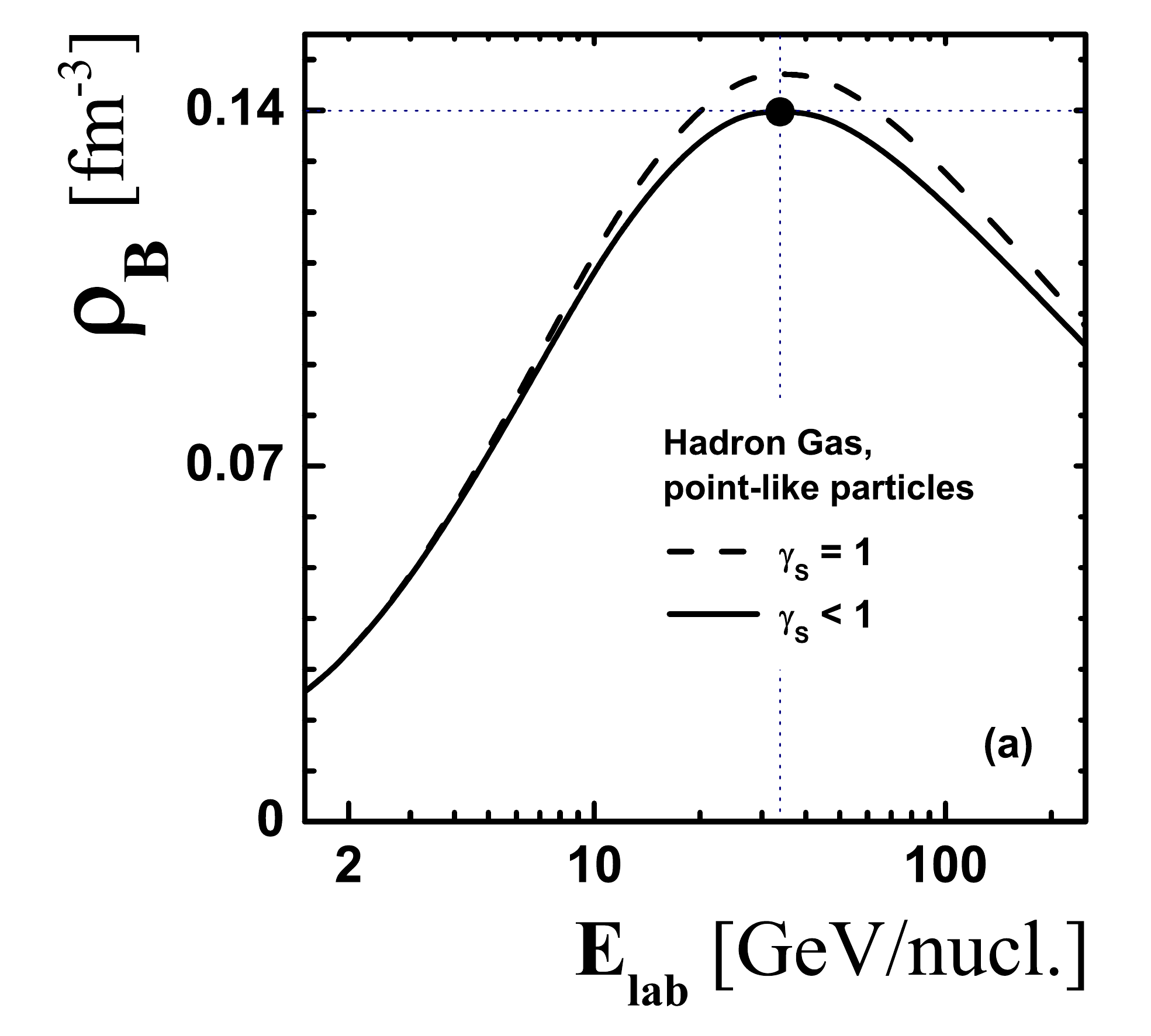,width=0.49\textwidth}~~
 \epsfig{file=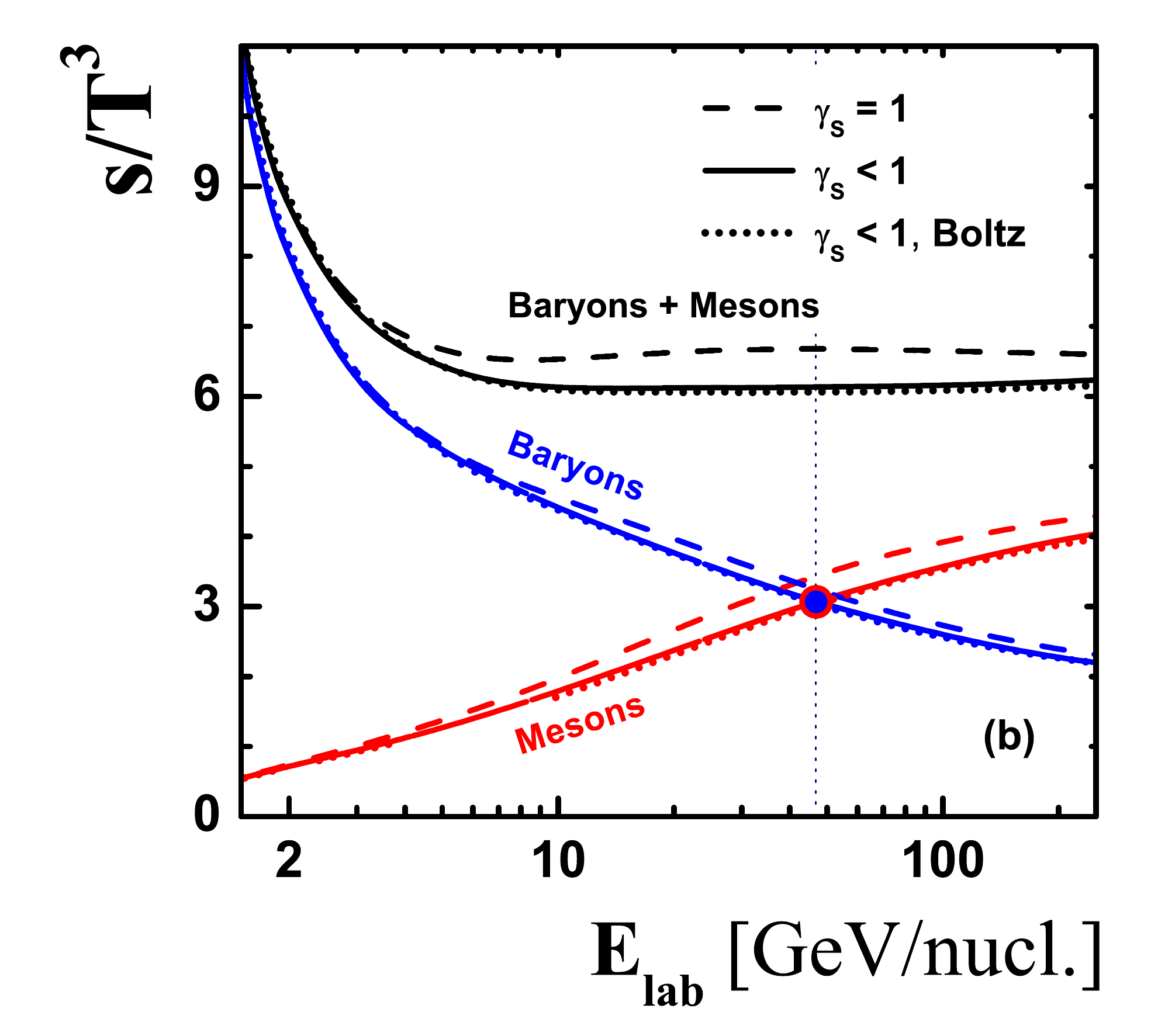,width=0.49\textwidth}
 \caption{(Color online)
 The (a) net-baryon density $\rho_B$ and (b) the ratio
 $s/T^3$ along the chemical freeze-out line Eq.~(\ref{T}) and
 $\gamma_S$ according to Eq.~(\ref{gamma-S})  are shown by the solid lines. The dashed lines
correspond to $\gamma_S=1$ for (a) $\rho_B$ and (b) $s/T^3$. The
dotted line corresponds to the Boltzmann approximation in
$s/T^3$.}\label{fig-rhoB}
 \end{center}
\end{figure}

The net-baryon density $\rho_B^{id}$  as a function of
collision energy calculated following the freeze-out line
(\ref{T}) is shown in Fig.~\ref{fig-rhoB}~(a).
In this and the following figures the laboratory collision energy
per projectile nucleon $E_{lab}$ is used to present the dependence
on collision energy.  Its connection to center-of-mass energy per
nucleon pair, $\sqrt{s_{NN}}$, is given by
% \cite{PDG2008}:
% \eq{
$ \sqrt{s_{NN}} = \sqrt{2m_N\,E_{lab}+ 2m_N^2} $, where $m_N$ is
the nucleon mass.
As seen in Fig.~\ref{fig-rhoB} (a), the net-baryon density
has a maximum~\cite{RC2006} at  $E_{lab}\cong 34A$~GeV. This is
the collision energy at which the NA49 Collaboration observed the
maximum of the $K^+/\pi^+$ ratio (the {\it horn}) and other
signals of the onset of deconfinement~\cite{Horn}.

The total entropy density as a function of  collision energy
following the freeze-out  line Eq.~(\ref{T}) is shown in
Fig.~\ref{fig-rhoB}~(b).
Meson $s_{M}$ and baryon $s_{B}$ entropy densities are also
presented in the figure. With increasing collision energy the
baryon-dominated ($s_B>s_M$) matter changes to meson-dominated
($s_M>s_B$) matter. For the I--HRG model this transition is
located at $E_{lab}\cong 46A$~GeV.

For the  $T$-$\mu_B$ values at the chemical freeze-out line
Eq.~(\ref{T}) the role of quantum statistics   is small. For
baryons  the Fermi statistics  changes their densities by less
than 1\%. The largest density change due to the Bose statistics is
for pions. It is, however, still smaller than 10\%.  The ratio
$s/T^3$ calculated within the Boltzmann approximation, i.e.
$\eta=0$ in Eq.~(\ref{pres-id}), is shown in Fig.~\ref{fig-rhoB}
(b) by the dotted line. The deviations from the results with
quantum statistics included are hardly visible. They are even
smaller for $\rho_B$ and thus the corresponding dotted line
calculated with the Boltzmann approximation is not plotted.
The collision energy dependence of $\rho_B$ and $s/T^3$
calculated for the $\gamma_S$ parametrization Eq.~(\ref{gamma-S}) and
for $\gamma_S=1$ is also shown in Fig.~\ref{fig-rhoB}.
One concludes that the energy at which $\rho_B$ has the maximum as
well as the energy of the transition between baryon-dominated and
meson-dominated matter are approximately independent of the
quantum statistics and the degree of strangeness equilibration.

It was suggested \cite{s-bar} that the maximum of the net-baryon
density and/or the transition from  baryon to meson dominance may
be related to the anomalous behavior of the $K^+/\pi^+$
ratio~\cite{Horn}. In the next section these phenomena are
examined by taking into account the repulsive interactions between
hadrons.

\section{Excluded Volume Hadron--Resonance Gas}\label{sec-vdw}
The results presented in Section~II have been obtained within the
ideal hadron-resonance gas model in which only attractive
interactions between hadrons are taken into account by the inclusion of
resonances. In this section the role of repulsive interactions is
considered within the excluded volume hadron-resonance gas model.

The van der Waals excluded volume procedure corresponds to a
substitution of the system volume $V$ by the available volume $V_{{\rm av}}$,
\eq{ V~\rightarrow~V_{{\rm av}}~=~V~-~\sum_{i} v_i N_i~,\label{Vav}
}
where
$ v_i~=~4 \cdot \left(4\pi r^3_i/3\right)$
is the excluded volume parameter and $r_i$ is the corresponding
hard sphere radius of a particle $i$. This result, in particular,
the presence of a factor of 4 in the expression for $v_i$, can be
rigorously obtained for a low density gas of particles of a single
type (see, e.g., Ref.~\cite{Landau}).
In the grand canonical ensemble, the substitution (\ref{Vav}) %$V\rightarrow V-\sum_i v_iN_i$
leads to a transcendental equation for the pressure of the
EV--HRG\footnote{ A discussion of other excluded volume
formulations can be found in Ref.~\cite{vdw1}.} \cite{vdw,vdwa}:
\begin{eqnarray}\label{pres-vdw}
p~=~\sum_{i}p_i^{id}(T,\tilde{\mu_i})~;~~~~~ \tilde{\mu_i} ~=~
\mu_i ~- ~v_i\,p ~.
\end{eqnarray}
%
%The particle number density reads:
%
%\eq{\label{ni}
%
%n_i ~=~ \frac{\partial p}{\partial \mu_i}
% ~=~ \frac{n_i^{id}(T,\tilde{\mu_i})}{1+\sum_j v_j n_j^{id}(T,\tilde{\mu_j})}~.
%
%}
%
Using Eq.~(\ref{rhoB}) one finds the net-baryon, entropy and
energy densities:
%
%}
%
\eq{\label{dens}
 \rho_B ~=~ \frac{\sum_i b_i\,n_i^{id}
 (T,\tilde{\mu_i})}{1+\sum_j v_j n_j^{id}(T,\tilde{\mu_j})}~,&&
 s ~=~ \frac{\sum_i s_i^{id}(T,\tilde{\mu_i})}
 {1+\sum_j v_j n_j^{id}(T,\tilde{\mu_j})}~,&&
 \varepsilon ~=~ \frac{\sum_i\varepsilon_i^{id}(T,\tilde{\mu_i})}
 {1+\sum_j v_j n_j^{id}(T,\tilde{\mu_j})}~.
}
In comparison to the corresponding densities calculated within the
I--HRG model the densities in the EV-HRG model Eq.~(\ref{dens}) are lower
because of two reasons:

i) due to the suppression factor $[1+\sum_j
v_jn_j^{id}(T,\tilde{\mu_j})]^{-1}$~~~and

ii) due to the shift in chemical potential
$\mu_i\rightarrow\tilde{\mu_i}$ which in the Boltzmann
approximation leads to the suppression factor $\exp(-v_ip/T)<1$.

The shift of the chemical potential makes the Boltzmann
approximation even more accurate than in the case of  the ideal
gas. If all proper volume parameters are the same $v_i=v$ (i.e.
$r_i=r$), the Boltzmann approximation gives the total suppression
factor $R$
\eq{\label{R}
 R(T,\mu_B;~r)
 ~=~ \frac{\exp\left(-~v\,p/T\right)}{1+v\sum_j
 n_j^{id}(T,\tilde{\mu_j})}~,
% \langle N_i\rangle
% ~=~ R\; V\; n_i^{id}(T,\mu_B).
%
}
the same for all densities of Eq.~(\ref{dens}):
\eq{\label{dens-vdw}
\rho_B(T,\mu_B)=R\,\rho_B^{id}(T,\mu_B)~,~~~
s(T,\mu_B)=R\,s^{id}(T,\mu_B)~,~~~
\varepsilon(T,\mu_B)=R\,\varepsilon^{id}(T,\mu_B)
}
 and
%
%\eq{\label{sup-dens}
%
$n_i(T,\mu_B)= R\,n_i^{id}(T,\mu_B)$~.
Typical values of hard-core radii considered in the
literature~\cite{st2,Hama-2004,Werner-Karpenko-2010,HRG-vs-Lattice,radii}
are $r=(0.3\div 0.8)$~fm.

\begin{figure}[ht!]
\begin{center}
 \epsfig{file=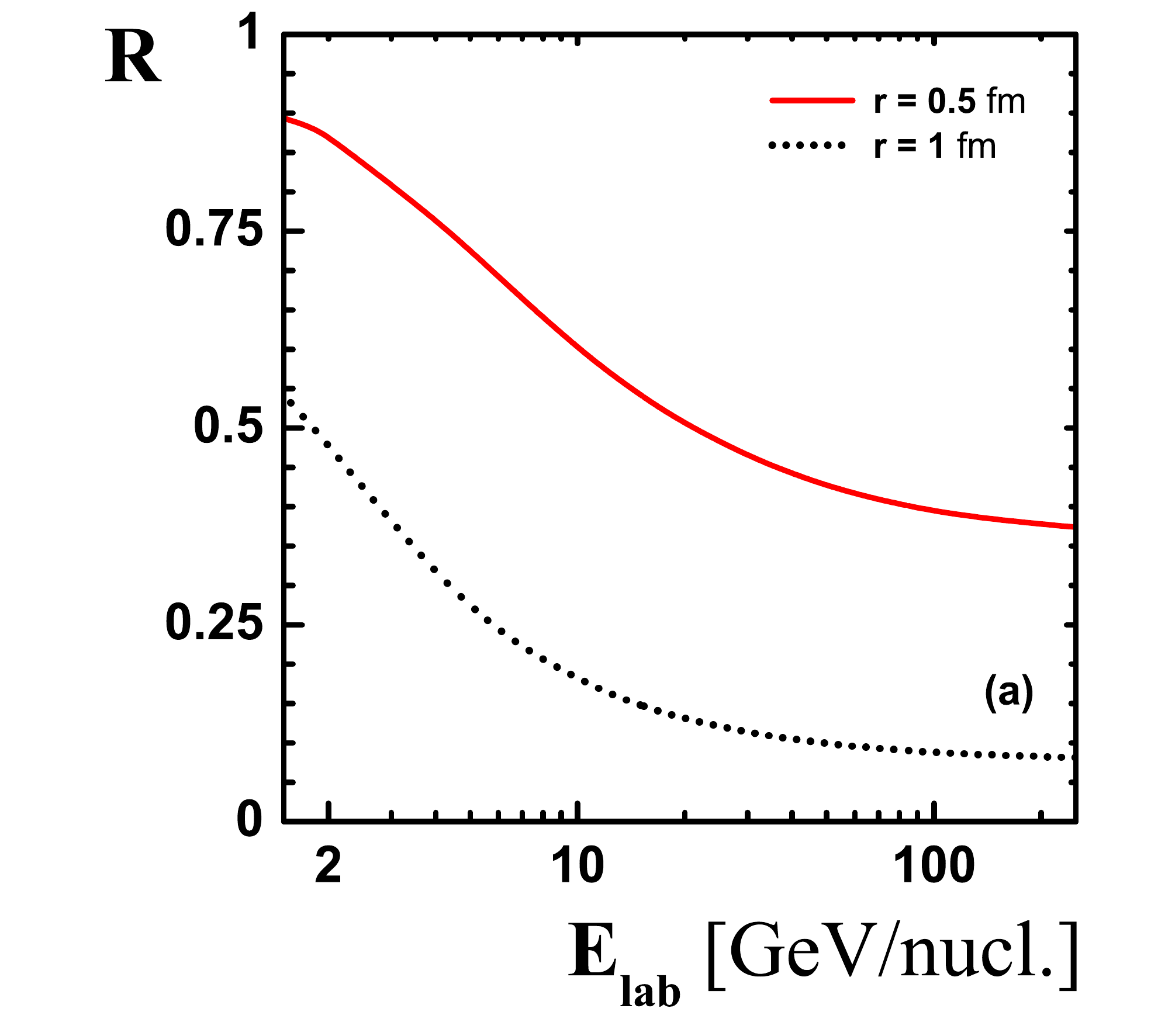,width=0.49\textwidth}~~
 \epsfig{file=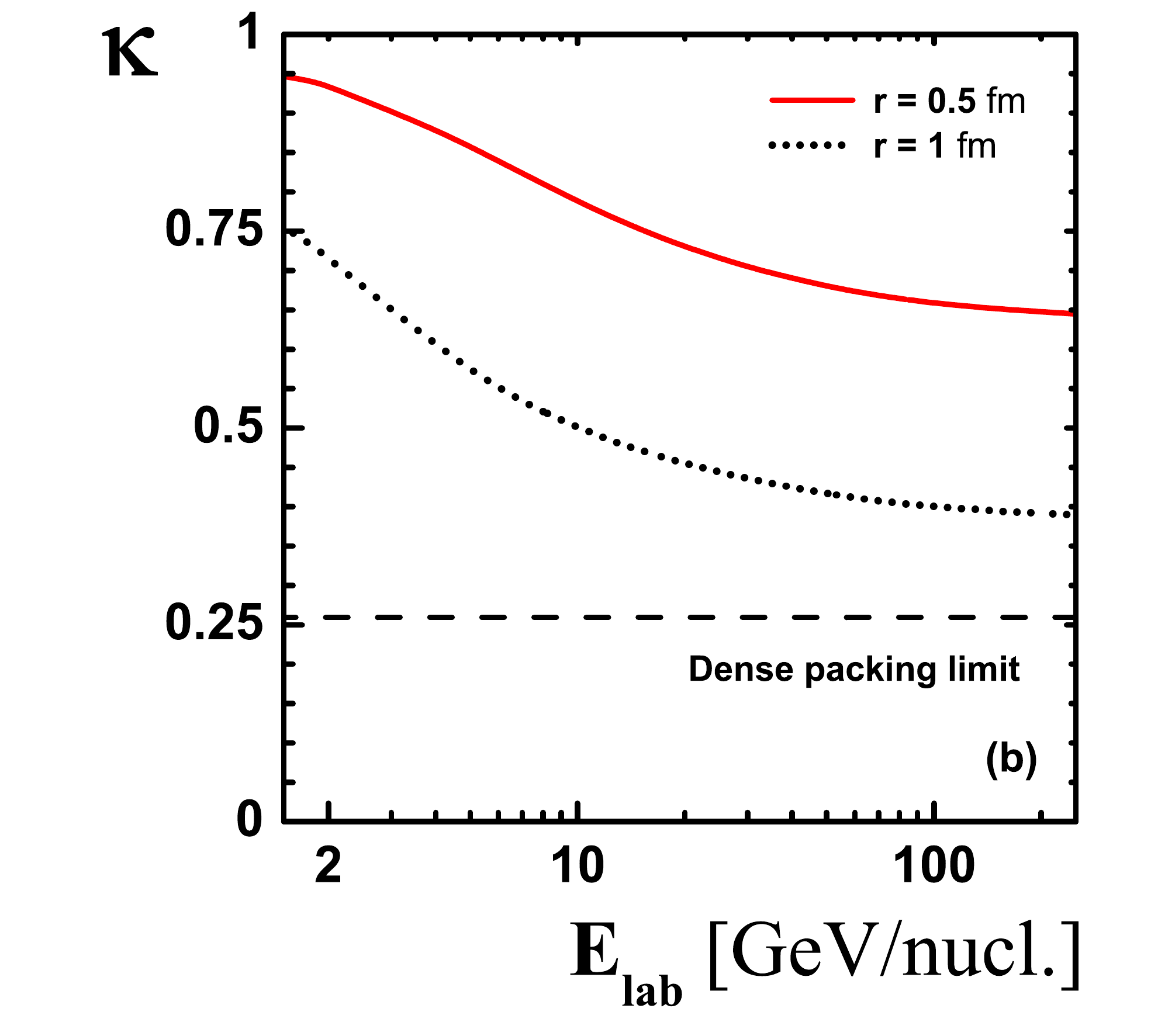,width=0.49\textwidth}
 \caption{(Color online) (a) The excluded volume suppression factor
 $R$ Eq.~(\ref{R}) and (b) the fraction of the available volume $\kappa$
 Eq.~(\ref{frac})
  as functions of $E_{lab}$ along the chemical freeze-out line
  Eqs.~(\ref{T}) and (\ref{gamma-S}). The solid and dotted  lines correspond
  to $r=0.5$~fm and $r=1$~fm, respectively. The dashed line in (b)
  corresponds to the dense packing limit 0.26 for hard spheres. }\label{fig-R}
 \end{center}
\end{figure}

\begin{figure}[ht!]
\begin{center}
 \epsfig{file=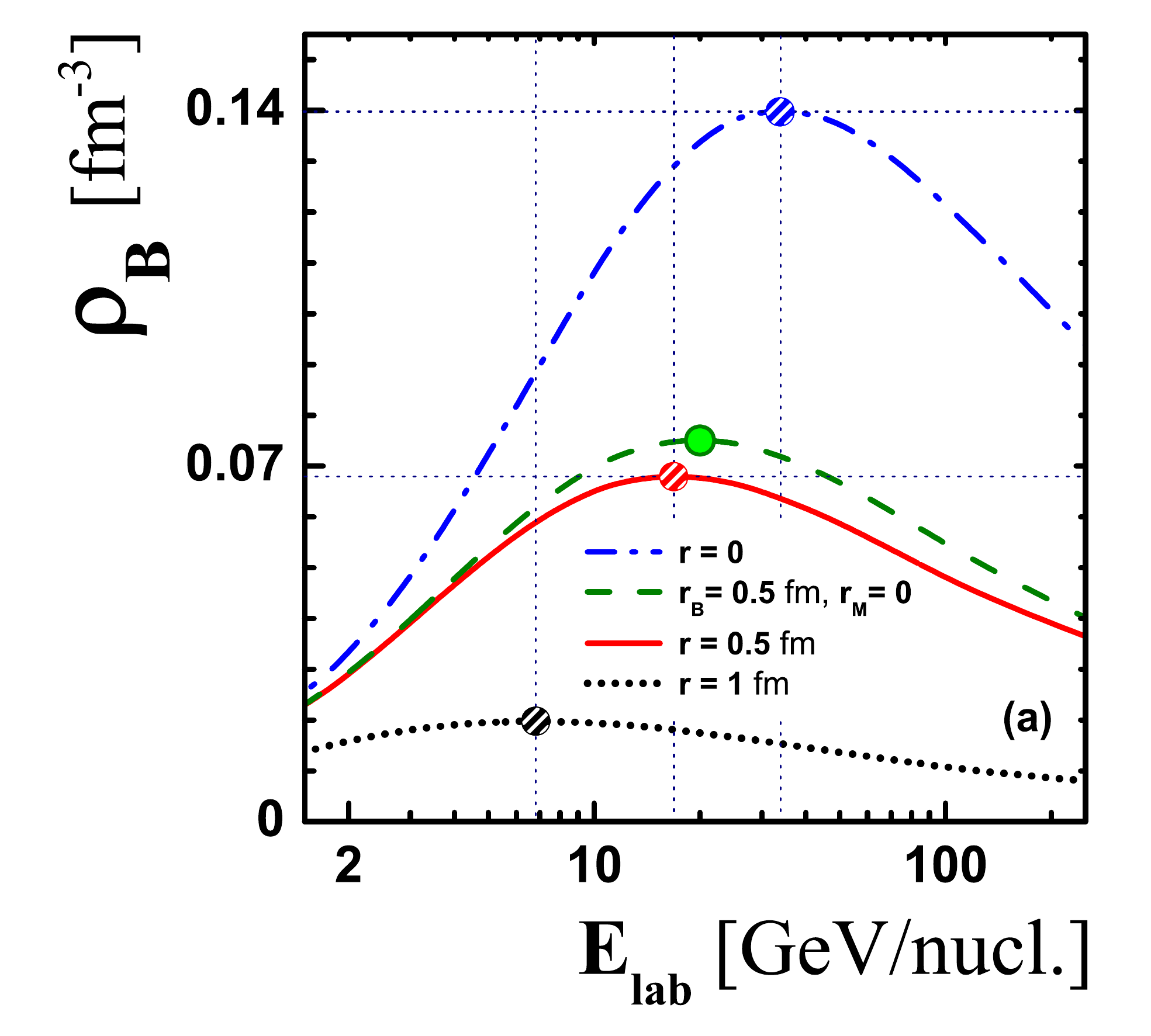,width=0.49\textwidth}~~
 \epsfig{file=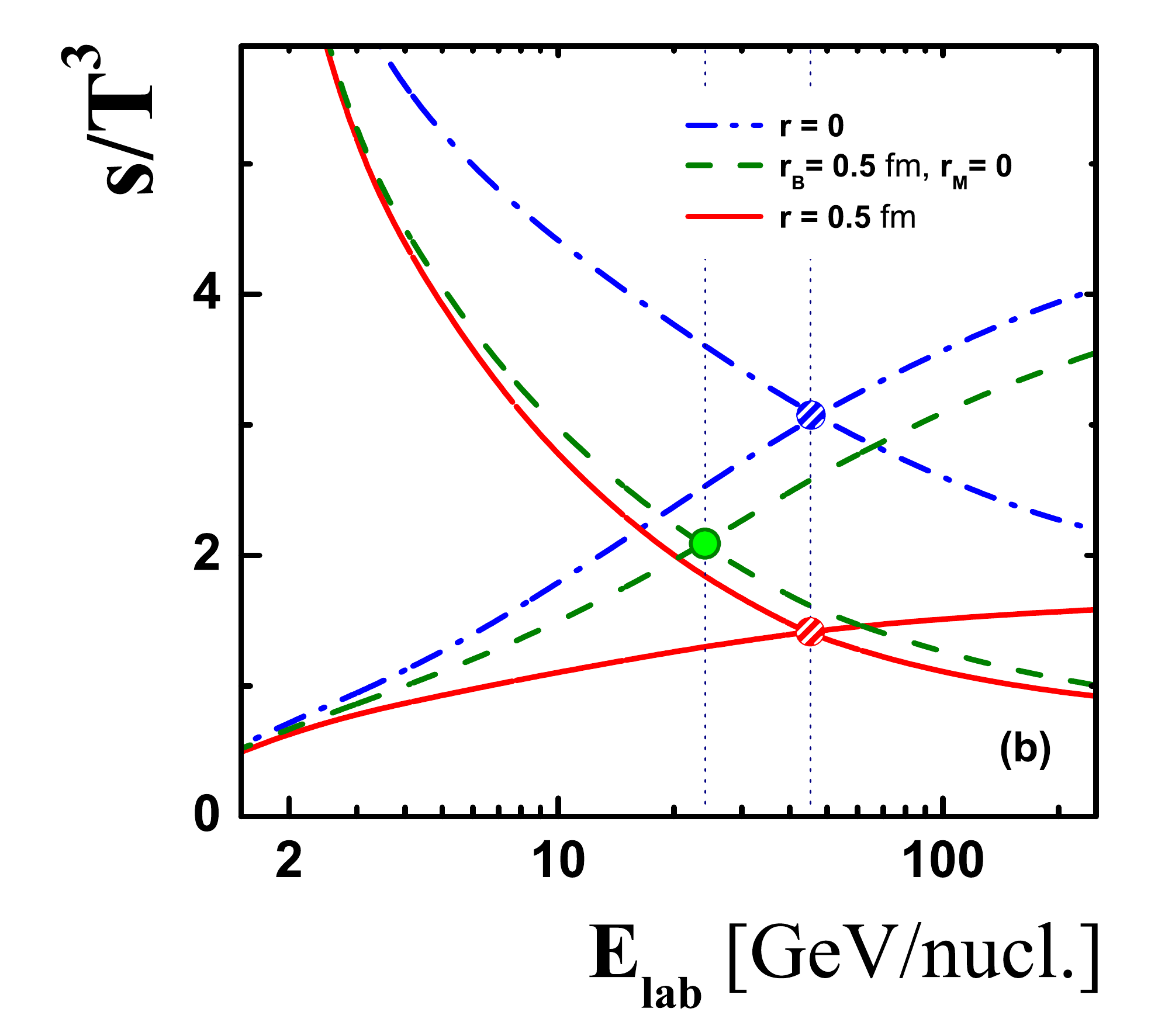,width=0.49\textwidth}
 \caption{(Color online) (a): The net-baryon density along the chemical
 freeze-out line from Eqs.~(\ref{T}) and (\ref{gamma-S}). Dashed-dotted line
 corresponds to the model with $r=0$, dashed line to
 $r_B=0.5$~fm and $r_M=0$, solid line to $r_B=r_M=r=0.5$~fm,
 and dotted line to $r_B=r_M=r=1$~fm. (b) The ratios $s_B/T^3$ and
 $s_M/T^3$  along the chemical
 freeze-out line Eqs.~(\ref{T}) and (\ref{gamma-S}). Dashed-dotted lines
 correspond to the model with $r=0$, solid lines to
 $r_B=r_M=r=0.5$~fm, and dashed-dotted lines
 to $r_B=0.5$~fm and $r_M=0$.
} \label{fig-rhoB-vdw}
 \end{center}
\end{figure}

The energy dependence of the suppression factor Eq.~(\ref{R})
calculated along the chemical freeze-out line for $r=0.5$~fm and
$r=1$~fm  is shown in Fig.~\ref{fig-R}~(a).
The $R$ factor (\ref{R}) decreases monotonously with increasing
collision energy. For example, for $r=0.5$~fm one finds $R\cong
0.9$ and $R\cong 0.4$ at small and large $E_{lab}$, respectively.
One may therefore expect a decrease of the value of  $\rho_B$ at
its maximum by a factor of 0.5 (for $r=0.5$~fm), and a shift of
the position of the maximum to a smaller collision energy. In
fact, in Fig.~\ref{fig-rhoB-vdw}~(a) one observes that the maximum
of the net-baryon density is located at $E_{lab}\cong 17A$~GeV for
$r=0.5$~fm and at $E_{lab}\cong 7A$~GeV for $r=1$~fm, instead of
$E_{lab}\cong 34A$~GeV for the I--HRG model. It is also seen that
the value of $\rho_B$ at the maximum decreases strongly with the
increasing value of the hard-core radius. The entropy density
shown in Fig.~\ref{fig-rhoB-vdw} (b) is reduced by the same
suppression factor. The collision energy at which the baryon and
meson entropy densities are equal is, however, independent of $R$
and is located at $E_{lab}\cong 46A$~GeV. This is however true
only if the hard-core radius $r$ is the same for all hadrons.

A fraction of the total volume $\kappa\equiv V_{{\rm av}}/V$,
which is available for the extended hadrons, can be estimated as
follows for equal baryon and meson radiuses:
\eq{\label{frac}
 \kappa
 %\equiv \frac{V_{{\rm av}}}{V}
 ~=~\frac{V-v\sum_i N_i}{V}
 =1-v\sum_i n_i(T,\tilde{\mu_i})
 =1-\frac{v~\sum_i n_i^{id}(T,\tilde{\mu_i})}{1+v\sum_jn_j^{id}(T,\tilde{\mu_j})}
 =\exp\left(\frac{v\,p}{T}\right)\,R~.
 }
The parameter $\kappa$ is shown in Fig.~\ref{fig-R}~(b) for
$r=0.5$ and $r=1$~fm. One can see that $\kappa$ is always larger
than  the dense packing limit for hard spheres:
$1-\pi/(3\sqrt{2})\cong 0.26$~\cite{Landau}. This ensures a
consistency of the excluded volume approach at all collision
energies even for the largest considered radius $r=1$~fm. We also
remind that the excluded volume parameter $v$ is assumed to be
four times larger than the hadron volume $4\pi r^3/3$.

It is  interesting %instructive
to consider the role of the excluded volume effects for different
hard-core radii of baryons $r_B$ and mesons $r_M$. As an example,
the results for $r_B=0.5$~fm and $r_M=0$ are presented in
Fig.~\ref{fig-rhoB-vdw}. In a comparison to the results for
$r=0.5$~fm for all hadrons one observes small changes of $\rho_B$
but a significant shift of the transition point between the baryon
and meson dominated matter. Its position decreases from
$E_{lab}\cong 46A$~GeV to $E_{lab}\cong 23A$~GeV.

The model with non-equal hard-core radii ($r_B=0.7$~fm and $r_M=0$) was
already used in Ref.~\cite{Hama-2004}. The EV--HRG
models with non-equal radii for different
hadron species  need, however, further detailed studies. This is because
fits to the hadron yields performed with the EV--HRG model with
non-equal radii give different freeze-out parameters $T$
and $\mu_B$ than those in Eq.~(\ref{T}) obtained within the I--HRG
model.

\begin{figure}[ht!]
\begin{center}
 \epsfig{file=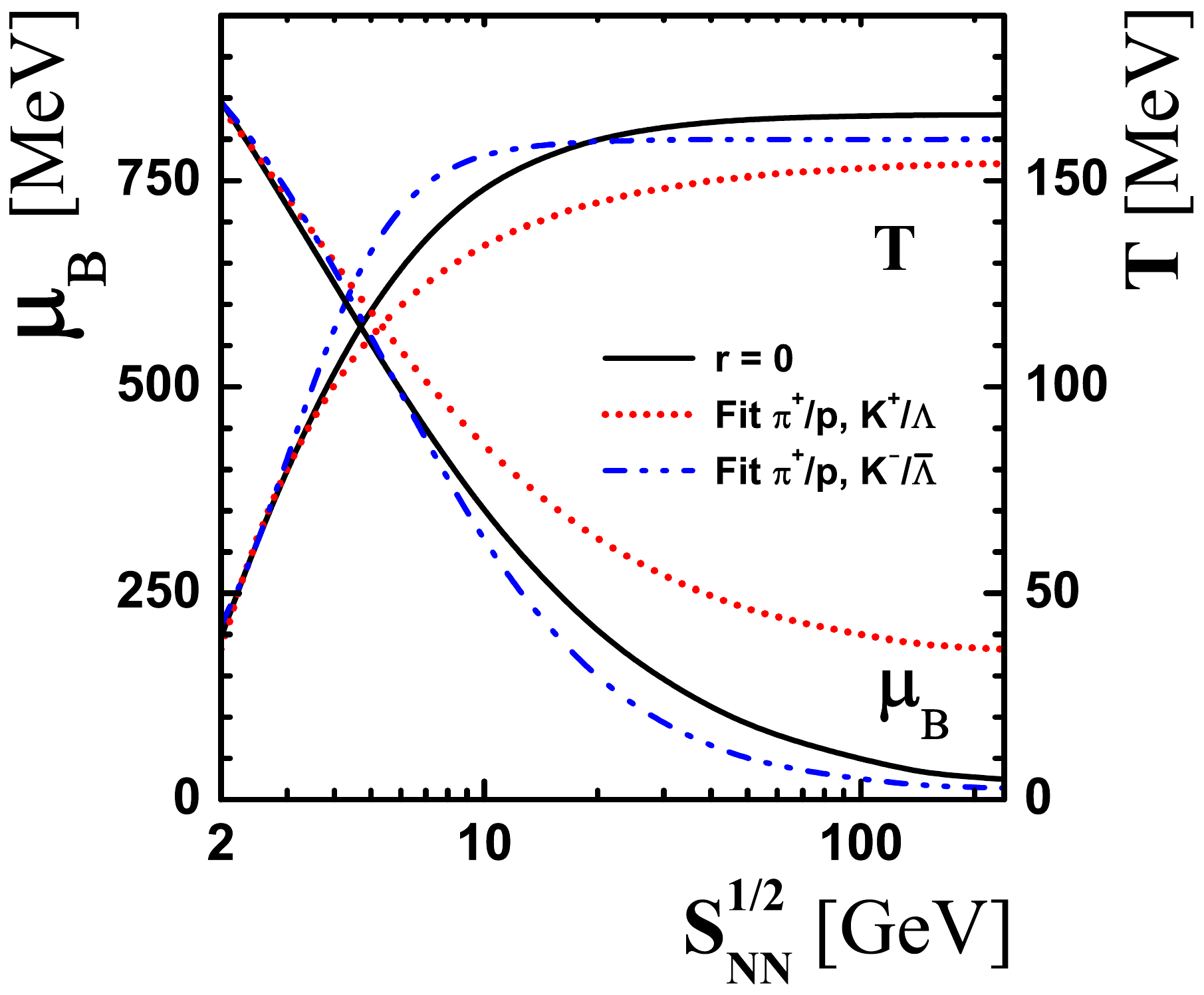,width=0.49\textwidth}
 \caption{(Color online)
 Comparison of the freeze-out line~(\ref{T}) (solid line), and the lines obtained using
 the $\pi^+/p$, $K^+/\Lambda$ (dotted line) and $\pi^+/p$,
 $K^-/\overline{\Lambda}$ ratios (dash-dotted line) within the
 EV--HRG model with $r_B=0.5>r_M=0$~fm.
 The ratios are calculated within I--HRG along the freeze-out line, Eq.~(\ref{T}), see text for details.
 } \label{fig-Freeze-out-vdw}
 \end{center}
\end{figure}

As an illustration, we estimate the possible changes of $T$ and
$\mu_B$ freeze-out parameters due to the excluded volume effects
with $r_B\neq r_M$.
At least two particle ratios are required to determine $T$ and
$\mu_B$. In the presented examples these ratios are calculated
using the parameters at the freeze-out line (\ref{T}) obtained for
$r_B=r_M$. Then using these ratios new freeze-out parameters $T$
and $\mu_B$ are calculated within the EV--HRG model with
$r_B=0.5$~fm and $r_M=0$. First,  the pion to proton, $\pi^+/p$,
and kaon to lambda, $K^+/\Lambda$, ratios are selected. These
mesons and baryons are the most abundant particles. Second the
$\pi^+/p$ and $K^-/\overline{\Lambda}$ ratios, which includes
antibaryon, are considered. The results are shown in
Fig.~\ref{fig-Freeze-out-vdw}.

The new 'freeze-out lines' significantly deviate
from the one obtained within the I-HRG model, Eq.~(\ref{T}).
These deviations are also strongly dependent on the ratios or
multiplicities selected for the analysis.
For different reactions different hadron sets are measured with
different precision.
Therefore, accurate estimates of $r_B$ and $r_M$ from the data on
hadron multiplicities would require a significant dedicated effort.

\begin{figure}[ht!]
\begin{center}
 \epsfig{file=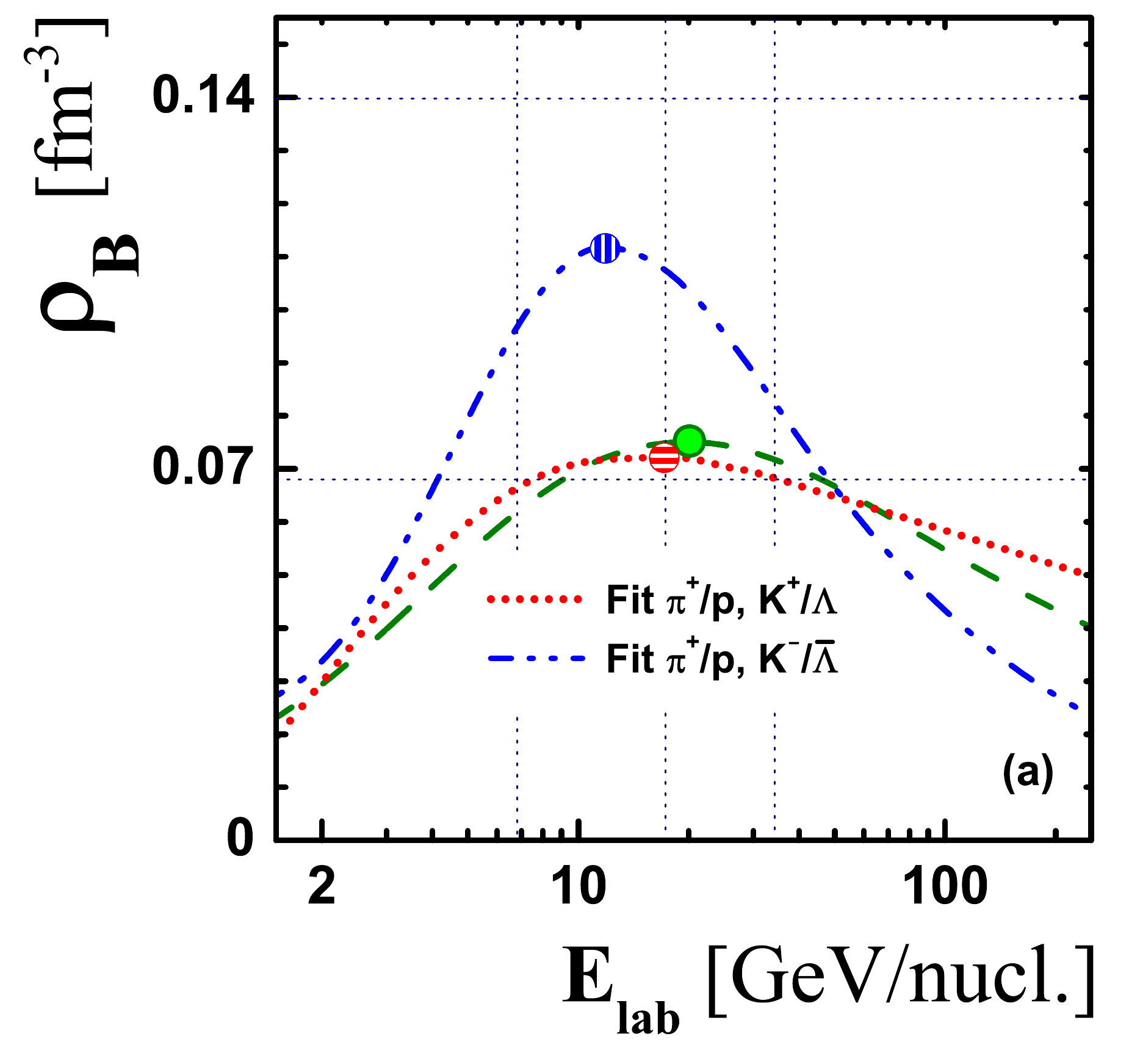,width=0.49\textwidth}~~
 \epsfig{file=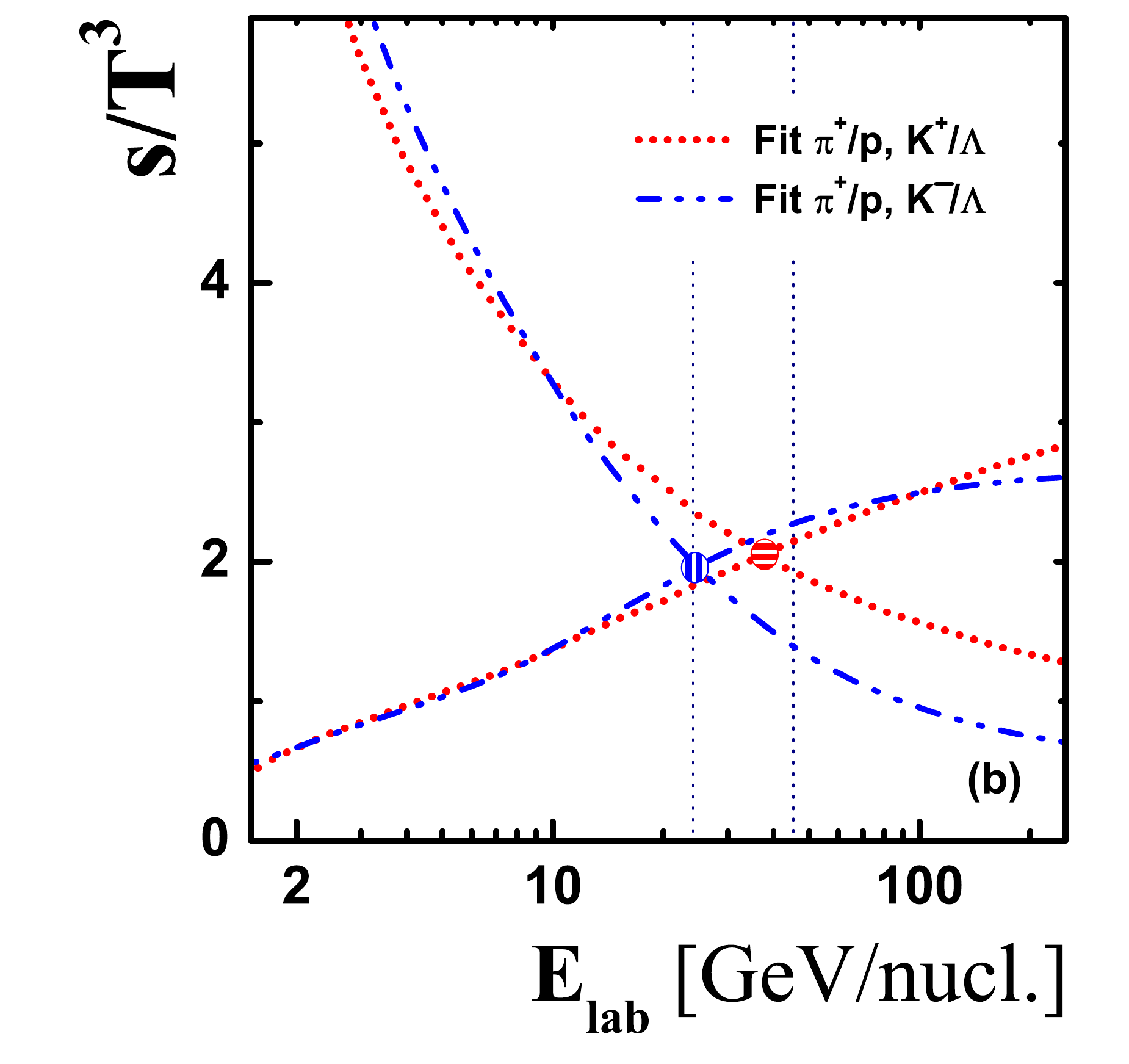,width=0.49\textwidth}
 \caption{(Color online)
 The same as in Fig.~\ref{fig-rhoB-vdw} for EV--HRG with
 $r_B=0.5>r_M=0$~fm. Dotted line corresponds to the fit of $\pi^+/p$, $K^+/\Lambda$
 and dash-dotted line to the fit of the $\pi^+/p$,
 $K^-/\overline{\Lambda}$ ratios. Dashed line, vertical and horizontal dotted lines are the same as in
 Fig.~\ref{fig-rhoB-vdw}.
 } \label{fig-Transition-Points-vdw}
 \end{center}
\end{figure}

The baryon number density and baryon/meson entropy densities along
the new freeze-out lines from Figs.~\ref{fig-Freeze-out-vdw} are shown in
Fig.~\ref{fig-Transition-Points-vdw}.
The new fits with $r_B=0.5$~fm and $r_M=0$ change the details but
preserve the main features of the system with $r_B=r_M=0.5$~fm.
In particular, the position of the net-baryon density maximum depends basically
on the $r_B$ parameter while the position of the baryon/meson transition point is
sensitive to the difference between the $r_B$ and $r_M$ parameters.

Particle number fluctuations are straightforwardly sensitive to
the hard-core hadron radius $r_B=r_M$ \cite{fluc}.  In a recent
paper~\cite{JFu} the same freeze-out line as well as the THERMUS
program has been used for the analysis of the event-by-event
particle number fluctuations. Higher moments of the net-proton
multiplicity distribution were calculated  and compared with the
STAR data. The results suggest that the hadron hard-core radius
$r_B=r_M$ is in the range from $0.3$~fm to $0.5$~fm. However, for
the final conclusion the important effects of the exact charge
conservation \cite{SM-vs-Exp} and the experimental acceptance
\cite{Bzdak-Koch} should be also included and their consequences
within the EV--HRG model should be studied.

\section{Summary}\label{sec-sum}

The ideal hadron-resonance gas model is simple and has only a few
free parameters. In spite of this it is successful in describing
the bulk properties of mean hadron multiplicities in high energy
collisions. The model takes into account attractive interactions
between hadrons via a presence of resonances, but ignores
repulsive interactions. The repulsive interactions  are, however,
needed to catch the basic qualitative features of strong
interactions, e.g, the phase transition between hadron-resonance
gas and the quark-gluon plasma: point-like hadrons and resonances
would be a dominant phase at very high energy densities as their
total degeneracy factor is much larger than that of quarks and
gluons. Moreover, the repulsive interactions strongly modify the
properties of the hadron-resonance gas. The most common way to
include repulsive interactions in the hadron-resonance gas model
is to follow the van der Waals excluded volume procedure and
introduce the hard-core radii of hadrons.

If radii of all hadrons are assumed to be the same, the chemical
freeze-out parameters, temperature and baryon chemical potential,
fitted to data on mean hadron multiplicities are identical to
those obtained within the ideal hadron-resonance gas model.
However, all densities calculated within the van der Waals model
are lower than the corresponding densities obtained within the
ideal gas model and thus the fitted volume parameter in the van
der Waals  gas formulation is significantly larger. The density
suppression factor $R$ depends on the $T$ and $\mu_B$ parameters,
which in turn depend on collision energy. Consequently, the
collision energy dependence of densities is sensitive to the
assumed hard-core radius of hadrons. In particular, the energy at
which net-baryon density has a maximum decreases from about
$E_{lab} \cong 34A$~GeV for the ideal gas model to about $E_{lab}
\cong 7A$~GeV for the excluded volume model with $r = 1$~fm.
If the radii of hadrons are assumed to be different, the densities
of different hadrons are modified differently. Clearly, the
excluded volume effects are even larger for the hadron matter  at
stages preceding the chemical freeze-out in nucleus-nucleus
collisions, i.e. at larger values of the energy density.

In view of these studies, the estimates of collision energies at
which the net-baryon density at the chemical freeze-out reaches
its maximum and/or the transition between baryon and meson
dominated matter takes place are premature. One needs to renew a
search for a suitable set of the excluded volume parameters.
Experimental and/or theoretical methods to better estimate
hard-core radii of hadrons within the excluded volume model are
needed to improve our understanding of the properties of
hadron-resonance matter. If all hard-core radiuses are equal to
each other, the particle number ratios are not  sensitive to their
numerical value. Thus, the data on average multiplicities are not
enough and independent measurements of the total system volume is
needed. However, the particle number fluctuations depend
straightforwardly on the hard-core hadron radius \cite{fluc}.
Precise measurements of higher moments of hadron multiplicity
distribution in nucleus-nucleus collisions are now in progress. An
interpretation of these data within the EV-HRG opens the way to
estimate the value of hard-core radius $r$ from the data.
%

%%%%%%%%%%%%%%%%%%%%%%%%%%%%%%%%%%%%%%%%%%%%%%%%%%%%%%%%%%%%%%%%%%%%

\begin{acknowledgments} We are thankful to
W.~Greiner,  M.~Hauer, Iu.~Karpenko, L.M.~Satarov, P.~Seyboth, and
V.~Voronyuk for fruitful discussions. This work was supported by
the Humboldt Foundation, by the Program of Fundamental Research of
the Department of Physics and Astronomy of NAS, Ukraine, and by
the German Research Foundation under Grant No. GA 1480/2-1 and the
HICforFAIR Grant No. 20130403.

\end{acknowledgments}


\begin{thebibliography}{100}

%\bibitem{PDG2008}
%  C.~Amsler {\it et al.}  [Particle Data Group Collaboration],
%  %``Review of Particle Physics,''
%  Phys.\ Lett.\ B {\bf 667}, 1 (2008).

\bibitem{st1}
J. Cleymans and H. Satz, Z. Phys. C {\bf 57}, 145 (1993).

\bibitem{st2}
  G.~D.~Yen, M.~I.~Gorenstein, W.~Greiner, and S.~N.~Yang,
  %``Excluded volume hadron gas model for particle number ratios in A+A collisions,''
  Phys.\ Rev.\ C {\bf 56}, 2210 (1997);
  G.~D.~Yen and M. I. Gorenstein, Phys. Rev. C {\bf 59}, 2788
  (1999).

\bibitem{st3}
  F.~Becattini, J.~Cleymans, A.~Keranen, E.~Suhonen and K.~Redlich,
  %``Features of particle multiplicities and strangeness production in central heavy ion collisions between 1.7A-GeV/c and 158A-GeV/c,''
  Phys.\ Rev.\ C {\bf 64}, 024901 (2001).
% [hep-ph/0002267].
%
% F. Becattini, J. Cleymans, J.C. Ker\"anen, E.~Suhonen, and
% K.~Redlich, Phys. Rev. C {\bf 64}, 042901 (2001).

\bibitem{st3a}
P. Braun-Munzinger, D. Magestro, K.~Redlich, and J.~Stachel, Phys.
Lett. B {\bf 518}, 41 (2001).

\bibitem{st3b}
J. Rafelski and J. Letessier, Nucl. Phys. A {\bf 715}, 98c (2003).

\bibitem{st4}
A. Andronic, P. Braun-Munzinger, and J. Stachel, Nucl. Phys. A
{\bf 772}, 167 (2006).

\bibitem{Gaz2005}
  F.~Becattini, J.~Manninen and M.~Gazdzicki,
  %``Energy and system size dependence of chemical freeze-out in relativistic nuclear collisions,''
  Phys.\ Rev.\ C {\bf 73}, 044905 (2006).
% [hep-ph/0511092].
%
% J.~Manninen,  F.~Becattini,  and M.~Ga\'zdzicki,
%  %``Energy and system size dependence of chemical freeze-out in relativistic nuclear collisions,''
%  Phys.\ Rev.\ C {\bf 73}, 044905 (2006).
%  %  [hep-ph/0511092].


\bibitem{Rafelski}
 %\gamma_s
 J.~Rafelski, Phys.\ Lett.\ {\bf B262}, 333 (1991); P.~Koch, B.~Muller, J.~Rafelski, Phys.\ Rep.\ {\bf 142}, 167 (1986).
 %
 J.~Letessier, J.~Rafelski, A.~Tounsi, Phys.\ Rev.\ {\bf C50}, 406 (1994);
 C.~Slotta, J.~Sollfrank, U.~Heinz, AIP Conf.\ Proc.\ (Woodbury) {\bf 340}, 462 (1995).
%
%\gamma_q
 J.~Letessier and J.~Rafelski,
 %``Chemical non-equilibrium and deconfinement in 200-A/GeV sulphur induced reactions,''
 Phys.\ Rev.\ C {\bf 59}, 947 (1999)  [hep-ph/9806386].

\bibitem{Becattini}
  F.~Becattini and U.~W.~Heinz,
  %``Thermal hadron production in p p and p anti-p collisions,''
  Z.\ Phys.\ C {\bf 76}, 269 (1997)  [Erratum-ibid.\ C {\bf 76}, 578 (1997)]
  [hep-ph/9702274];
%
  F.~Becattini,
  %``A Thermodynamical approach to hadron production in e+ e- collisions,''
  Z.\ Phys.\ C {\bf 69}, 485 (1996).

\bibitem{LHC}
 M.~Floris,
 %``Identified particles in pp and Pb-Pb collisions at LHC energies with the ALICE detector,''
 J.\ Phys.\ G {\bf 38}, 124025 (2011)  [arXiv:1108.3257 [hep-ex]];
%
 A~Kalweit [ALICE Collaboration],
 %GLOBAL PROPERTIES OF STRANGE PARTICLE PRODUCTION IN pp AND Pb?Pb COLLISIONS WITH THE ALICE DETECTOR
 %Presented at the Conference ?Strangeness in Quark Matter 2011?, Krak?w, Poland, September 18?24, 2011.
 Acta\ Phys. Polon.\ B, Proc.\ Suppl.\ {\bf 5}, 225 (2012);
%
 B.~Abelev {\it et al.}  [ALICE Collaboration],
 %``J/psi production at low transverse momentum in Pb-Pb collisions at sqrt(sNN) = 2.76 TeV,''
 arXiv:1202.1383 [hep-ex].


\bibitem{DM}
 R.~Dashen, S.-K.~Ma, H.J.~Bernstein, Phys. Rev. {\bf 187}, 345 (1969);
 R.~Dashen, S.-K.~Ma, Phys. Rev. A {\bf 4}, 700 (1971).

\bibitem{vdw}
M.~I. Gorenstein, V.~K. Petrov, and G.~M. Zinovjev, Phys. Lett. B
{\bf 106}, 327 (1981);
%
%\bibitem{vdw}
D.~H.~Rischke, M.~I. Gorenstein, H. St\"ocker, and W. Greiner, Z.
Phys. C {\bf 51}, 485 (1991).
%

\bibitem{Wal}
B.~D.~Serot and J.~D.~ Walecka, {\it Advances in Nuclear Physics}
(Plenum, New York, 1986), Vol 16; Int. Journ. Mod. Phys. E {\bf
6}, 515 (1997).

\bibitem{RMF}
O. Lourenco, M. Dutra, A. Delfino, and M. Malheiro,  Phys. Rev. D
{\bf 84}, 125034 (2011).


\bibitem{vdwa}
J. Cleymans, M.~I.~Gorenstein, J. Stalnacke, and E. Suhonen, Phys.
Scripta {\bf 48}, 277 (1993).

\bibitem{vdw2}
  M.~I.~Gorenstein, H.~Stoecker, G.~D.~Yen, S.~N.~Yang, and W.~Greiner,
  %``Chemical freezeout in relativistic A+A collisions: Is it close to the QGP?,''
  J.\ Phys.\  G {\bf 24}, 1777 (1998).
  %  [nucl-th/9711055].

\bibitem{Hama-2004}
  Y.~Hama, T.~Kodama and O.~Socolowski, Jr.,
  %``Topics on hydrodynamic model of nucleus-nucleus collisions,''
  Braz.\ J.\ Phys.\  {\bf 35}, 24 (2005)  [hep-ph/0407264].

\bibitem{Werner-Karpenko-2010}
  K.~Werner, Iu.~Karpenko, T.~Pierog, M.~Bleicher and K.~Mikhailov,
  %``Event-by-Event Simulation of the Three-Dimensional Hydrodynamic Evolution from
  % Flux Tube Initial Conditions in Ultrarelativistic Heavy Ion Collisions,''
  Phys.\ Rev.\ C {\bf 82}, 044904 (2010)  [arXiv:1004.0805 [nucl-th]].

  \bibitem{Sa}
   L.~M. Satarov, M.~N. Dmitriev, and I.~N. Mishustin,
   %Equation of state of
 %hadron resonance gas and the phase diagram of strongly interacting matter,
  Phys. Atom. Nucl. {\bf 72}, 1390 (2009);
 A.~V. Merdeev, L.~M. Satarov, and I.~N. Mishustin,
 %Hydrodynamic modeling of deconfinement phase transition in heavy-ion collisions
%at NICA-FAIR energy domain,
Phys. Rev. C {\bf 84}, 014907 (2011).


\bibitem{HRG-vs-Lattice}
  A.~Andronic, P.~Braun-Munzinger, J.~Stachel and M.~Winn,
  %``Interacting hadron resonance gas meets lattice QCD,''
  arXiv:1201.0693 [nucl-th].

%\bibitem{T-Hagedorn}
% R. Hagedorn, CERN-TH-4100/85 (1985).

\bibitem{RC2006}
J. Randrup and J. Cleymans, Phys. Rev. C {\bf 74}, 047901 (2006).

\bibitem{s-bar}
  J.~Cleymans, H.~Oeschler, K.~Redlich, and S.~Wheaton,
  %``Comparison of chemical freeze-out criteria in heavy-ion collisions,''
  Phys.\ Lett.\ B {\bf 615}, 50 (2005).

\bibitem{Cleymans2005}
  J.~Cleymans, H.~Oeschler, K.~Redlich, and S.~Wheaton,
   Phys.\ Rev.\ C {\bf 73}, 034905 (2006).

\bibitem{Cleymans2011}
J. Cleymans, Phys. Part. Nucl. Lett., Vol. {\bf 8}, No. 8, 797
(2011).


\bibitem{Horn}
  M.~Ga\'zdzicki and M.~I.~Gorenstein,
  %``On the early stage of nucleus-nucleus collisions,''
  Acta Phys.\ Polon.\  B {\bf 30}, 2705 (1999);
 % [arXiv:hep-ph/9803462];
%
  S.~V.~Afanasiev {\it et al.}  [The NA49 Collaboration],
  %``Energy dependence of pion and kaon production in central Pb + Pb
  %collisions,''
  Phys.\ Rev.\  C {\bf 66}, 054902 (2002);
%  [arXiv:nucl-ex/0205002];
%
  C.~Alt {\it et al.}  [NA49 Collaboration],
  %``Pion and kaon production in central Pb + Pb collisions at 20-A
  %and 30-A-GeV: Evidence for the onset of deconfinement,''
  Phys.\ Rev.\ C {\bf 77}, 024903 (2008);
  %  [arXiv:0710.0118 [nucl-ex]].
 M.~Ga\'zdzicki, M.~Gorenstein and P.~Seyboth,
  %``Onset of deconfinement in nucleus-nucleus collisions:
  % Review for pedestrians and experts,''
  Acta Phys.\ Polon.\ B {\bf 42}, 307 (2011).


%\bibitem{raf}
%J. Rafelski, Phys. Lett. B {\bf 62}, 333 (1991).

\bibitem{NA61}
%\cite{Aduszkiewicz:2012fr}
  A.~Aduszkiewicz {\it et al.} [NA61 Collaboration],
  %``NA61/SHINE at the CERN SPS: Plans, status and first results,''
  Acta Phys.\ Polon.\ B {\bf 43}, 635 (2012)
  [arXiv:1201.5879 [nucl-ex]].
  %%CITATION = ARXIV:1201.5879;%%

\bibitem{BES}
%\cite{Aduszkiewicz:2012fr}
  G.~Odyniec {\it et al.} [STAR Collaboration],
  Acta Phys.\ Polon.\ B {\bf 43}, 627 (2012).

\bibitem{NICA}
  A.~N.~Sissakian {\it et al.}  [NICA Collaboration],
  %``The nuclotron-based ion collider facility (NICA) at JINR: New prospects for heavy ion collisions and spin physics,''
  J.\ Phys.\ G {\bf 36}, 064069 (2009);
  A.~Sorin, V.~Kekelidze, A.~Kovalenko, R.~Lednicky, I.~Meshkov and G.~Trubnikov,
  ``Heavy-ion program at NICA/MPD at JINR,''  Nucl.\ Phys.\ A {\bf 855}, 510 (2011).

\bibitem{CBM}
  B.~Friman, (ed.), C.~Hohne, (ed.), J.~Knoll, (ed.), S.~Leupold, (ed.), J.~Randrup, (ed.),
  R.~Rapp, (ed.) and P.~Senger, (ed.),
  %``The CBM physics book: Compressed baryonic matter in laboratory experiments,''
  Lect.\ Notes Phys.\  {\bf 814}, 1 (2011).

\bibitem{FAIR}
  H.~Stoecker and C.~Sturm,
  %``The FAIR start,''
  Nucl.\ Phys.\ A {\bf 855}, 506 (2011).

\bibitem{Galatyuk}
 T.~Galatyuk,
 {\it Investigation of baryon rich dense nuclear matter at SIS100},
 %Presentation given on the
 "CPOD-2013", % - 8th International Workshop on Critical Point and Onset of Deconfinement",
 March 11-15, 2013, Napa, California, USA
 [https://www-alt.gsi.de/documents/DOC-2013-Mar-43-1.pdf].

\bibitem{THERMUS}
  S.~Wheaton and J.~Cleymans,
  %``THERMUS: A Thermal model package for ROOT,''
  Comput.\ Phys.\ Commun.\  {\bf 180}, 84 (2009)  [hep-ph/0407174].

\bibitem{Landau}
 L.~D.~Landau and E.~M.~Lifshitz, Statistical Physics (Oxford:
Pergamon) 1975.

\bibitem{vdw1}
M.~I. Gorenstein, Phys. Rev. C {\bf 86}, 044907 (2012).
% arXiv:1205.1762 [nucl-th].

\bibitem{radii}
  P.~Braun-Munzinger, K.~Redlich and J.~Stachel,
  %``Particle production in heavy ion collisions,''
  In *Hwa, R.C. (ed.) et al.: Quark gluon plasma* 491-599  [nucl-th/0304013]; %and references therein.
%
  P.~Braun-Munzinger, I.~Heppe and J.~Stachel,
  %``Chemical equilibration in Pb + Pb collisions at the SPS,''
  Phys.\ Lett.\ B {\bf 465}, 15 (1999)  [nucl-th/9903010].

%
\bibitem{fluc}
M. I. Gorenstein, M. Hauer, and D. O. Nikolajenko, Phys. Rev. C {\bf 76},
024901 (2007).


\bibitem{JFu}
  J.~Fu,
  %``Higher moments of net-proton multiplicity distributions in heavy ion collisions at chemical freeze-out,''
  Phys.\ Lett.\ B {\bf 722}, 144 (2013).

\bibitem{SM-vs-Exp}
  V.~V.~Begun, M.~Gazdzicki, M.~I.~Gorenstein and O.~S.~Zozulya,
  %``Particle number fluctuations in canonical ensemble,''
  Phys.\ Rev.\ C {\bf 70}, 034901 (2004)  [nucl-th/0404056];
  %
  V.~V.~Begun, M.~Gazdzicki, M.~I.~Gorenstein, M.~Hauer, V.~P.~Konchakovski and B.~Lungwitz,
  %``Multiplicity fluctuations in relativistic nuclear collisions: Statistical model versus experimental data,''
  Phys.\ Rev.\ C {\bf 76}, 024902 (2007)  [nucl-th/0611075].

\bibitem{Bzdak-Koch}
  A.~Bzdak, V.~Koch and V.~Skokov,
  %``Baryon number conservation and the cumulants of the net proton distribution,''
  Phys.\ Rev.\ C {\bf 87}, 014901 (2013)  [arXiv:1203.4529 [hep-ph]];  %%CITATION = ARXIV:1203.4529;%%
%
  A.~Bzdak and V.~Koch,
  %``Acceptance corrections to net baryon and net charge cumulants,''
  Phys.\ Rev.\ C {\bf 86}, 044904 (2012)  [arXiv:1206.4286 [nucl-th]].


%  D.~R.~Oliinychenko, K.~A.~Bugaev and A.~S.~Sorin,
  %``Investigation of hadron multiplicities and hadron yield ratios in heavy ion collisions,''
%  arXiv:1204.0103 [hep-ph].

%\bibitem{Bohr}
% A. Bohr and B. Mottelson, "Nuclear Structure", Vol.1, Benjamin, New York, (1969).


%\bibitem{hard-core}
%hard-core radius

%\bibitem{PDG}
%  K.~Hagiwara {\it et al.}  [Particle Data Group Collaboration],
  %``Review of particle physics. Particle Data Group,''
%  Phys.\ Rev.\ D {\bf 66}, 010001 (2002).



\end{thebibliography}
\end{document}